

\def\Winf{{\cal W}_{1+\infty}}
\def\bb{{\bf b}}
\def\cb{{\bf c}}
\def\ZZ{{\bf Z}}
\def\vep{\varepsilon}
\def\gen{{\cal P}}
\def\KacFormula{Kac-Radul Formula\ }

\input harvmac
\Title{\vbox{\baselineskip12pt\hbox{UT-661}\hbox{
}%
}}{\vbox{\centerline{Free Fields and}
	\vskip2pt\centerline{Quasi-Finite
Representation of $\Winf$ Algebra}}}

\centerline{Yutaka MATSUO\footnote{$^\dagger$}
{E-mail address: matsuo@tkyux.phys.s.u-tokyo.ac.jp}}
\bigskip\centerline{Department of Physics,
University of Tokyo}\centerline{Bunkyo-ku, Hongo 7-3-1}
\centerline{Tokyo 113, Japan}

\vskip .3in
We study quasi-finite representation of the
$\Winf$ algebra recently proposed by Kac and Radul.
When the central charge is integer, we show
 that they are represented by
free fermions and bosonic ghosts.
There are some nontrivial representations
with vanishing central charge.
We discuss that they may be described by
large $N$ limit of topological models.
We calculate their operator algebras explicitly.

\Date{December 1993}

\newsec{Introduction}
The quantum extension of $\Winf$ algebra appears in many places
in the two dimensional physics.
In quantum gravity, it comes from
Schwinger-Dyson equation of the matrix
model\ref\rFKN{M.~Fukuma, H.~Kawai, R.~Nakayama
{\sl Comm. Math. Phys.} {\bf 143} (1992) 371-403.}.
 Similar structure
appears in the discrete states of quantum
gravity coupled with $c=1$
matter\ref\rKP{I.~Klebanov, A.M.~Polyakov,
{\sl Mod. Phys. Lett} {\bf A6} (1991) 3273
\semi
A.~Dhar, G.~Mandal, S.~Wadia, ``Non-relativistic Fermions,
Coadjoint Orbits of $W_{1+\infty}$ algebra and
String Field Theory at $c=1$, Tata Institute Preprint,
TIFR-TH-92/40 (June 1992).},
although the relation between those two is not yet quite clear.
In quantum hole effect, it appears as the realization of
magnetic translation group
\ref\rC{A.~Cappeli, C.~Trugenberger, G.~Zemba,
{\sl Nucl. Phys.} {\bf B396} (1993) 49;
preprint MPI-Ph/93-75, DFTT 65/93 (1993) hep-th/9310181.}.
\ref\rKIS{S.~Iso, D.~Karabali, B.~Sakita,
{\sl Phys. Lett.} {\bf B296} (1992) 143.}
Also in turbulence in two dimensions,
it might play important r\^ole since the fluid motion in
two dimension defines the area-preserving diffeomorphism.

So far, however, the representation theory of $\Winf$ algebra
\ref\rB{I.~Bakas, {\sl Phys. Lett.} {\bf B228} (1989) 57.}
\ref\rPRS{ C.N.~Pope, X.~Shen,  L.J.~Romans, {\sl Nucl.Phys.}
 {\bf 339B} (1990) 191.}
was not understood well
compared with other loop algebras such as
Virasoro algebra or Kac-Moody algebra.
Only free fermion representation
\ref\rDJKM{E.~Date, M.~Jimbo, M.~Kashiwara and T.~Miwa,
in {\sl Proceedings of RIMS Symposium on Nonlinear Integral Systems},
Kyoto, 1981, edited by M.~Jimbo and T.~Miwa (World Scientific, Singapore,
1983)}
\ref\rBK{I.~Bakas and E.~Kiritsis,
{\sl Nucl.Phys} {\bf B343} (1990) 185.}
\ref\rO{S.~Odake, {\sl Int. J. of Mod. Phys.} {\bf A7}
(1992) 6339.} is studied to a certain extent.
Although
the appearance of $\Winf$ algebra itself is interesting,
we do not know how to extract information from the symmetry.

$\Winf$ algebra
may be characterized by its infinite number of generator fields,
$W_1$, $W_2$, $W_3$, $W_4\ldots$.  Consequently, for each energy level,
we may have infinite number of independent states.  In such situation, it is
a non-trivial question when we have finite number
of non-vanishing states at each energy level
(quasi-finite representation).
Recently, Kac and Radul \ref\rKR{V. Kac and A.Radul,
MIT Mathematics preprint, July 1993, hep-th/9308153.} gave
a convincing answer to this problem.  They have
derived the possible form of the highest weights of
$\Winf$ module. They have also clarified
the condition when the representation becomes unitary.

In this letter, we give an explicit realization of their work
in terms of free fields.
Roughly speaking, the representations of $\Winf$ algebra
can be classified into two groups by the value of the central charge,
(1) $C\neq 0$ and (2) $C=0$.  In the first case
with further condition that $C=$ Integer, we show that
they are essentially given by combining fermionic and bosonic ghosts.
We give an explicit proof of the dimension formula and
the null field conditions.
In the second case, we propose
that it may be described by the large
N limit of topological combination
of free fields. Finally, we derive the operator algebra
predicted by our construction.

\newsec{Brief Sketch of \KacFormula}
%
%
$\Winf$ algebra is
the  quantum version of the $w_{1+\infty}$ algebra generated by
polynomials of
$z$ and $D=z{\partial\over{\partial z}}$.
For any function $f(w)$ which is regular at $w=0$, we may write
its typical generator as $z^rf(D)$.
The two--cocycle which defines the quantization
is explicitly
given by,
\eqn\eCocy{\eqalign{&\Psi(z^r f(D), z^s g(D))=-\Psi(z^s g(D),
z^r f(D))\cr &= \cases{\sum_{1\leq j\leq r}f(-j)g(r-j)&
if $r=-s>0$\cr
0& if $r+s\neq 0$ or $r=s=0$}\cr}}
Denote the quantum generator
associated with classical operator $z^rf(D)$ as $W(z^rf(D))$.
$\Winf$ algebra may be written as,
%
%
\eqn\eWinf{\eqalign{&\left[ W(z^r f(D)), W(z^s g(D))\right]\cr=%
&W(z^{r+s}f(D+s)g(D))-W(z^{r+s}f(D)g(D+r))%
+ C \Psi(z^r f(D), z^s g(D)).\cr}}
It is easily derived by using a useful relation
$f(D)z^r=z^rf(D+r)$.

%
%

We use
$V^{n}_{k}=W(z^kD^n)$,
for generators in terms of mode.
$\Winf$ algebra in terms of
lower dimensional operators are given by
$$
\left[ V^0_r, V^0_s\right]=  Cr\delta_{r,-s},\qquad
\left[ V^1_r, V^1_s\right]= (s-r)V^1_{r+s}-
C{r^3-r\over 6}\delta_{r,-s}.
$$
We may understand that $V^0$ and $V^1$ stand for $U(1)$ current
and Virasoro generator respectively.
The central charge of
Virasoro algebra is given by $-2C$.
We may freely change it by
adding total derivative of $U(1)$ currents in the definition of
the Virasoro generator.

%
%
The highest weight state of the $\Winf$ algebra is defined by
following conditions,
\eqn\eHWC{\cases{V^n_r|\lambda>=0, & if $r>0$,\cr%
V^n_0|\lambda>=\Delta_n|\lambda>,& $n=0,1,2,\cdots,\infty$.}}
Needless to say, unlike other loop algebras,
we have infinite number of the weight vector $\Delta_n$.
For later convenience, we combine them
into the following form,
\eqn\eWF{\Delta(x)\equiv-\sum_{n=0}^\infty {x^n\over n!}\Delta_n.}
This is the eigenvalue of the operator, $W(-e^{xD})$.

%
%
The highest weight state $|\lambda>$ may be specified by
operators which annihilate it.  Those operators
form a subalgebra (parabolic subalgebra) of $\Winf$ algebra.
For each level $k$, we denote the set of its elements as $I_k$.
We may easily check that if $z^{-k}P(D)$ is an element of of $I_k$,
any element of the form $z^{-k}P(D)f(D)$ is again the element of
$I_k$.
It shows that $I_k$ is an ideal of the polynomial algebra.
We use $\gen_k(D)$ as the generator of ideal $I_k$.
For quasi-finite representation, $\gen_k(w)$ whould be a polynomial
of finite order with respect to $w$.

{\it The main observation of} \rKR\ {\it is
that if we determine the polynomial $\gen_1(D)\equiv \gen(D)$
at level 1, one can derive $\gen_k(D)$ for $k>1$
and level $0$ weight function, $\Delta(x)$.}

If the differences between each root of $\gen(D)$ are
not integer, level $k$ generator is given by,
\eqn\eLvlk{\gen_k(D)=\gen(D)\gen(D-1)\cdots \gen(D-k+1).}
On the other hand, the information at level $0$ may be
extracted by multiplying $W(zg(D))$ from left hand side to
the equation, $W(z^{-1}\gen(D)|\lambda>=0$.
By using commutation relation
\eWinf, one finds,
\eqn\eWPg{%
(W(\gen(D)g(D))-W(\gen(D+1)g(D+1))+C W(\gen(0)g(0)))|\lambda>=0.}
This relation holds for any $g(D)$ and gives enough information
on the highest weights $\Delta(x)$.
It is in particular useful to use the combination,
\eqn\eF{F(x)\equiv C+(1-e^x)\Delta(x)}
{}From \eWPg,
one may prove that it satisfies a differential equation,
$\gen({d\over {dx}})F(x)=0.$
If $\gen(w)$ is given by, $\gen(w)=(w-s_1)^{n_1}
\cdots(w-s_\ell)^{n_\ell}$,
the general solution is,
\eqn\eFG{F(x)=\sum_{i=1}^\ell p_{n_i}(x)e^{s_i x}.}
Here $p_n(x)$ is $n-1$ order polynomial in $x$.
Together with \eF, one gets the explicit form of the highest weights,
which is the main result of \rKR,
\eqn\eWFG{\Delta(x)={{\sum_{i=1}^\ell p_{n_i}(x)e^{s_i x}-C}\over%
{e^x-1}}.}
The central charge $C$ is determined by the requirement
that there is no singularity in \eWF\ when $x\rightarrow 0$.
If we decompose \eWFG, we realize that
there are basically two types of representations of $\Winf$ algebra,
\eqna\eWBasic
$$
\eqalignno{\Delta(x)=C{{e^{sx}-1}\over{e^x-1}},\qquad&
 C\neq 0.&\eWBasic a\cr
\Delta(x)={p_\ell x^\ell e^{sx}\over{e^x-1}},\qquad&
C=0,\ell=1,2,3\ldots &\eWBasic b\cr
}$$

The requirement of unitarity gives severe restriction on the polynomial
$p_n(x)$. It was proved \rKR\ that the weight function  should take the
following form,
$\Delta(x)=\sum_{i=1}^\ell n_i ({e^{s_\ell x}-1})/
(e^x-1)$,
with {\it positive integer} $n_i$
and {\it real $s_\ell$}.  The central charge
should be a positive integer,
$C=\sum_i^\ell n_i.$

\newsec{Free Field Realization}
In the following, we derive explicit realizations
that meets \KacFormula \eWFG.
Since we know free fermion
(or bosonic ghost) representation,
we should first check to which representation
they belong.
We describe free fermions and bosonic ghosts in
a parallel fashion.  We use the notation similar to
those in \ref\rFMS{D.~Friedan, E.~Martinec, S.~Shenker,
{\sl Nucl.Phys.} {\bf B271} (1986) 93.}.
Let $\bb(z)$ (resp. $\cb(z)$) stands for
either $b(z)$ or $\beta(z)$ (resp. $c(z)$ or $\gamma(z)$).
They are expanded as,
$\cb(z)=\sum_{\ell\in\ZZ}\cb_\ell z^{-\ell-s-1}$,
$\bb(z)=\sum_{\ell\in\ZZ}\bb_\ell z^{-\ell+s}$.
We use a general representation, i.e.
$s$ may not be restricted to be half integer. As a consequence,
$\cb$ and $\bb$
may not be invariant under $z\rightarrow ze^{2\pi i}$.
The (anti-) commutation relation is given by
$\left[ \cb_n, \bb_m\right]_\vep\equiv%
\cb_n\bb_m+\vep\bb_m\cb_n=\delta_{n+m,0}$,
with $\vep=1$ for fermions and $\vep=-1$ for bosons.
The vacuum $|s>$ is characterized by the conditions,
\eqn\eBCVac{\cb_\ell|s>=\bb_{\ell+1}|s>=0,
\qquad \ell=0,1,2,\ldots}

The generators of $\Winf$ algebra may be
described by those
$\bb$-$\cb$  fields by sandwiching.
For each classical generator of
the form, $z^rf(D)$, the corresponding
generator is defined by,
\eqn\eBCWgen{W(z^rf(D))=\vep\int{dz\over 2\pi i}:\cb(z)z^rf(D)
\bb(z):=\vep\sum_{\ell\in\ZZ}:\cb_{\ell}f(s+\ell)\bb_{-\ell}:.}
It is easy to see that this gives $\Winf$ algebra \eWinf\ with
$C=\vep$. The commutation relation with the free fields are given by,
\eqn\eBCW{
\left[W(z^rf(D)),\cb_\ell\right]=
\cb_{\ell+r}f(s+\ell),\quad
\left[W(z^rf(D)),\bb_\ell\right]=
-f(s-\ell-r)\bb_{\ell+s}.}

\vskip 5mm

\noindent{\bf Construction of Null States:}\hskip 3mm
As a first check that this free field construction
gives quasi-finite representation, let us examine the
parabolic subalgebra that annihilates the vacuum,
$W(z^{-k}\gen_k(D))|s>=0.$
We use the fact that a descendant state becomes null if it
satisfies the highest weight condition in terms of free fields
\eBCVac.

Let us first investigate the null-state at level 1.
Since,
$
\left[W(z^{-1}\gen(D)),\cb_{\ell}\right]\sim
\cb_{\ell-1},
$
the highest weight condition \eBCVac\ with $\ell=1,2,3\cdots$
holds trivially for any $\gen(D)$. For $\ell=0$, we need to impose,
$$
\left[W(z^{-1}\gen(D)),\cb_{0}\right]
=\cb_{-1}\gen(s)=0,\quad
\left[W(z^{-1}\gen(D)),\bb_{1}\right]
=-\bb_{0}\gen(s)=0.
$$
The minimum choice for $\gen(D)$ that satisfies those two
conditions is obviously,
\eqn\ebF{\gen(D)=D-s.}

The construction of null states at higher level
is similar.  For level $n$, nontrivial conditions arise
from $\ell=0,1,\cdots,n-1$.  Conditions,
$\gen_n(s+\ell)=0,$
for $\ell=0,1,\cdots,n-1$ should be satisfied.
The minimum polynomial
is \eqn\ebH{\gen_n(D)=(D-s)(D-s-1)\cdots(D-s-n+1).}
This computation manifestly confirm the general formula \eLvlk.
\vskip 5mm

\noindent{\bf Highest weights:}\hskip 3mm
Since we derive explicitly the generators of the
algebra and the null fields, the differential equation
shows that the highest weight should be given in the form,
$
\Delta(x)=n{e^{sx}-1\over e^x-1}.
$
with a free parameter $n$.
Actually, since $C=\vep$, $n$ should be
given by $n=\vep$.  Let us confirm it explicitly.

The $\Winf$ generator that have the eigenvalue
$\Delta(x)$ is given by,
\eqn\eWexD{%
W(-e^{xD})=
-\vep\int {dz\over 2\pi i}:\cb(z)e^{xD}\bb(z):=
-\vep\int {dz\over 2\pi i}:\cb(z)\bb(e^x z):.}
For the fermionic case, one may describe the vacuum $|s>$
through vertex operator,
\eqn\eFvac{|s>=e^{-s\phi(0)}|0>,\qquad
\partial_z\phi(z)=:b(z)c(z):}
$\phi(z)$ satisfies standard
OPE $\phi(z)\phi(0)\sim \log(z)$.
Using the vertex operator representation of free fermion,
$b(z)=e^{-\phi(z)}$, $c(z)=e^{\phi(z)}$, one may prove
the \KacFormula by a direct computation,
\eqn\ePrf{\eqalign{
&-\int{dz\over 2\pi i} (e^{\phi(z)}e^{-\phi(z e^x)})e^{-s\phi(0)}|0>\cr
&=-\int{dz\over 2\pi i}{(e^xz)^s\over z^s}{1\over %
{z-e^xz}}e^{-s\phi(0)}|0>={e^{sx}\over{e^x-1}}|s>.\cr}}
{}From this expression, we need to subtract ${1\over{e^x-1}}$
due to the normal ordering. This proves the formula \eWBasic{a}.
To derive a similar formula for the bosonic ghost, we need
to make use of the ``bosonization'' of bosoinic ghosts \rFMS,
$\beta=e^{-\sigma}\partial \xi,$
$\gamma=e^\sigma \eta,$
with $\xi(z)\eta(0)\sim {1\over z}$ and $\sigma(z)\sigma(0)
\sim -\log(z)$.
The vacuum $|s>$ is in this case represented as
$
|s>=e^{s\sigma(0)}|0>.
$
The proof of
the \KacFormula (with negative central charge $C=-1$)
can be done in the similar fashion as fermionic case.

Construction of the representation
for integer $C$ may be done by tensor products.
We remark that the minimal polynomial at level one
keeps its form $\gen(w)=w-s$ as long as we take the tensor product
of the same spin fields.
\vskip 5mm

\noindent{\bf Topological Representation:}\hskip 3mm
It is rather subtle to understand the second
type of representation, \eWBasic{b}.  In this case, the central
charge vanishes, suggesting that it may be related to
the topological field theory
\ref\rTFT{E. Witten,{\sl Nucl.Phys.}
{\bf B340} (1990) 281\semi
J.Distler, {\sl Nucl.Phys.} {\bf B342} (1990) 523.}.
Furthermore, the minimal
polynomial $\gen(w)$ need to have higher zeros $\gen(w)=(w-s)^n$ at
$w=s$. Since
a tensor product of finite free fields
gives only single zeros for $\gen(w)$,
we are forced to
realize them in limiting procedure.

We remark that \eWBasic{b}\ can be obtained from \eWBasic{a}\
by differentiation with respect to spin $s$ several times.
To take subtraction in the field theory,
one may combine
fields with different statistics \rTFT.
In our case, let us first combine
a pair of free fermion with spin $s'$
and a pair of bosonic ghost with spin $s$.
The \KacFormula becomes
$\Delta(x)={e^{xs'}-e^{xs}\over e^x-1}.$
The minimal polynomial for this system is
$\gen(w)=(w-s)(w-s')$.
Putting $s'=s+{p_1\over N}$ with one constant $p_1$, and
introduce $N$ replica of this $bc\beta\gamma$ system, one gets
$$\Delta(x)=N{e^{x(s+{p_1\over N})}-e^{xs}\over e^x-1}
={p_1xe^{sx}\over {e^x-1}}+O({1\over N}).$$
Since the minimal polynomial keeps its form
in the tensor product,
$$\gen(w)=\lim_{N\rightarrow\infty}(w-s-{p_1\over N})(w-s)
=(w-s)^2.$$
These are exactly conditions for \eWBasic{b} with $\ell=1$.

Higher representations $\ell>1$ may be obtained
similarly.  Let us
consider $M$ copies of
$m$ pairs of $bc\beta\gamma$ system with spins,
$(s_i, t_i)$ ($i=1\cdots m$).
The dimension formula in this case is,
\eqn\eDimOld{%
\Delta(x)={M \over {e^x-1}}\sum_{\ell=1}^m
\left(e^{s_\ell x}-e^{t_\ell x}\right) .}
Consider a case when all dimensions $(s_i, t_i)$ are
very close to a fixed value, $s$.  We rewrite each dimension as
$s_i=s+C_i\delta s$, $t_i=s+D_i \delta s$,
where  $C_i$ and $D_i$ are finite constants.
Define,
$
Q_\ell=\sum_{j=1}^m {{C_j^\ell-D_j^\ell}\over \ell !}.$
\eDimOld\ may then be expanded as follows,
\eqn\eDimN{%
\Delta(x)={Me^{sx}\over {e^x-1}}\sum_{\ell=1}^\infty
Q_\ell\delta s^\ell.}
In order to get $\ell$th order representation,
we need to require,
\eqn\eCond{Q_1=Q_2=\cdots=Q_{\ell-1}=0, \qquad
Q_\ell\neq 0.}
Replacing $M=N^\ell$ and
$\delta s={(p_\ell)^{1/\ell}\over NQ_\ell^{1/\ell}},$
one arrives at the desired answer,
$$
\lim_{N\rightarrow\infty}\Delta(x)={p_\ell x^\ell e^{sx}\over {e^x-1}}.
$$

\vskip 5mm

\noindent{\bf Operator Algebra: }\hskip 3mm
One merit to use free fields is that we do not need to worry about
consistency of the theory.  In general, it is a nontrivial
task to check the Jacobi identities of the
primary fields.  However, in our case,
it is satisfied automatically.

Computation of the operator algebra of
primary fields of type  \eWBasic{a} is
quite simple since it is identical to that of vertex operator.
Let us denote the primary field which corresponds to the dimension
formula as $\Phi_{\{n_i:s_i\}}(z)$.
One may explicitly represent it as the vertex operator
which generalize \eFvac.
The OPE between the vertex operators
is given by,
\eqn\eOPE{
\Phi_{\{n_i:s_i\}}(z)\cdot
\Phi_{\{n_i:s'_i\}}(w)\sim
(z-w)^{\sum_i n_i s_i s'_i}\Phi_{\{n_i:(s+s')_i\}}(w).}

On the other hand, we have to be much more cautious in
calculating OPE between the  primary fields of type \eWBasic{b}.
Let us denote primary field which corresponds to
\eWBasic{b} by $\Psi^{(\ell)}_{\{s:p_\ell\}}(z)$.
The precise form of the representation
is not unique. In other word, the primary field with the same index
$p_\ell$
may be written with different $C_\ell$ and $D_\ell$.
If we take OPE for vertex operators with
$(C,D)$ and $(C',D')$ which satisfy \eCond,
the composite vertex operator
also need to satisfy condition \eCond,
\eqn\eCondComp{
{Q''}_i\equiv \sum_{j=1}^m \left(
(C_j-C'_j)^i-(D_j-D'_j)^i\right)/i!=0,
\qquad i=0,1,\cdots,\ell-1.}
It is, however,  difficult to find general solutions
for these nonlinear equations.

In this letter, instead of using general
solution for the coefficients, we use
the simplest  solution for \eCondComp, i.e.
the two  vectors $(C,D)$ and $(C', D')$ should be proportional.
In this case, we may represent two primary fields
$\Psi^{(\ell)}_{\{s:p_\ell\}}$,
$\Psi^{(\ell)}_{\{s':{p'}_\ell\}}$  by
using the same coefficients $(C,D)$ but different $\delta s$.
Explicitly, $\delta s={{p_\ell}^{1/\ell}\over N{Q_\ell}^{1/\ell}}$
for the first operator,
and $\delta s'={{{p'}_\ell}^{1/\ell}\over N{Q_\ell}^{1/\ell}}$ for the
latter. After some algebra which generalize \eOPE, we obtain,
$$\Psi^{(\ell)}_{\{s:p_\ell\}}(z)
\Psi^{(\ell)}_{\{s':{p'}_\ell\}}(w)
\sim (z-w)^{\beta_\ell(s,s';p_\ell,{p'}_\ell)}
\Psi^{(\ell)}_{\{s+s':{p''}_\ell\}}(w).$$
where,
\eqn\ePBeta{{{p''}_\ell}^{1/\ell}=
{{p}_\ell}^{1/\ell}+{{p'}_\ell}^{1/\ell},\quad
\beta_\ell(s,s';p_\ell,{p'}_\ell)=\left\{
\eqalign{
&p_1 s'+{p'}_1 s \qquad(\ell=1)\cr
&2\sqrt{p_2{p'}_2}  \qquad(\ell=2)\cr
&0 \qquad(\ell>2)\cr}\right..}
The dependence on $N$ or $Q_\ell$ disappears
in these final relations, as it should be.
Vanishing of the exponents $\beta$ for $\ell>2$
cases is a simple consequence of the fact that
 $\Psi^{(\ell)}_{\{s:p_\ell\}}$ has dimension zero for these cases.

\newsec{Discussion}
This letter is only a
begining of our project and many
things should be clarified in the future.
First, one has to find the detail of the
Hilbert space, in particular, the
character formula\foot{We understand
that the character formula for $C=1$ case was
independently obtained by Awata, Fukuma, Odake and Quano
\ref\rAFOQ{H.~Awata, M.~Fukuma, S.~Odake and Y.~Quano,
``Eigensystem and Full Charactor Formula
of the $\Winf$ Algebra with $c=1$",
preprint YITP/K-1049, SULDP-1993-1 (December 1993).}.
We would like to thank M. Fukuma for explaining their
work.  It helps us to understand structure
of the null states.}
The character formula for \eWBasic{b}\ seems especially
interesting.

Secondly, it would be desirable to give representation
for $C=0$ by finite number of (possibly) interacting fields.
As far as using large $N$ limit, we have infinite number of
redundant degree of freedom. It may be problematic in the
future computation.  Such representation will be also
important in removing some arbitariness in the OPE computation
we have met.

Thirdly, we have to make connection with
the physical situation, such as quantum gravity
or quantum Hall effects. So far, only the unitary representation
was discussed.  However, as we have seen, $\Winf$ algebra
has much rich structure for the nonunitary cases.
Is there any place where such representation becomes useful?
Turbulence may be one of the candidate it
suffers the dispersion of enstrophy through viscocity.

\listrefs
\bye